\documentclass{aa}  

\usepackage{graphicx}
\usepackage{txfonts}
\usepackage{natbib}

\begin{document}

\title{Another forbidden solar oxygen abundance: the [\ion{O}{i}] 5577 \AA\ line}
\titlerunning{Another forbidden solar oxygen abundance}

\author{
J. Mel\'{e}ndez\inst{1,2} \and
M. Asplund\inst{3}
          }

%\offprints{J. Mel\'{e}ndez (\email{jorge@astro.up.pt}}

\institute{
Centro de Astrof\'{\i}sica da Universidade do Porto, Rua das Estrelas, 4150-762 Porto, Portugal \and
Research School of Astronomy \& Astrophysics, Australian National University, 
Mt. Stromlo Observatory, Weston ACT 2611, Australia \and
Max Planck Institute for Astrophysics,
Postfach 1317, 85741 Garching, Germany 
 }

\date{Received ...; accepted ...}

\abstract
% context heading (optional)
{Recent works with improved model atmospheres, line formation,
atomic and molecular data, and detailed treatment of blends, have 
resulted in a significant downward revision of the solar oxygen abundance.}
% aims heading (mandatory)
{Considering the importance of the Sun as an astrophysical standard and the
current conflict of standard solar models using the new solar abundances
with helioseismological observations 
%(e.g. Basu \& Antia 2008), 
we have performed a new study of the solar oxygen abundance 
based on the forbidden [\ion{O}{i}] line at 5577.34 \AA, not previously considered.}
% methods heading (mandatory)
{High-resolution (R $>$ 500 000), high signal-to-noise (S/N $>$ 1000) solar
spectra of the [\ion{O}{i}] 5577.34 \AA\ line have been analyzed employing both three-dimensional (3D)
and a variety of 1D (spatially and temporally averaged 3D, Holweger \& M\"uller, 
MARCS and Kurucz models with and without convective overshooting) model atmospheres.}
% results heading (mandatory)
{ The oxygen abundance obtained from the [\ion{O}{i}] 5577.3 \AA\ forbidden line is almost 
insensitive to the input model atmosphere and has a mean value of 
$\log \epsilon_{\rm O} = 8.71 \pm 0.02$ ($\sigma$ from using the different model 
atmospheres).
The total error (0.07 dex) is dominated by uncertainties in the log $gf$ value 
(0.03 dex), apparent line variation (0.04 dex) and uncertainties in the continuum
and line positions (0.05 dex).
}
% conclusions heading (optional) 
{
The here derived oxygen abundance is close to the 3D-based 
estimates from the two other [\ion{O}{i}] lines at
6300 and 6363\,\AA , the permitted \ion{O}{i} lines and vibrational and rotational
OH transitions in the infrared. Our study thus supports
a low solar oxygen abundance  ($\log \epsilon_{\rm O} \approx$ 8.7), 
independent of the adopted model atmosphere. 
}


\maketitle
%
%________________________________________________________________

\section{Introduction}

In recent years a significant reduction in the solar oxygen abundance has been proposed 
from $\log \epsilon_{\rm O} \equiv \log (N_{\rm O}/N_{\rm H}) + 12 = 8.93$
 (Lambert 1978; Sauval et al. 1984; Grevesse et al. 1984;
Anders \& Grevesse 1989)
down to  $\log \epsilon_{\rm O} \approx$ 8.7 or even lower
(Allende Prieto et al. 2001, 2004; Holweger 2001;
Asplund et al. 2004; Melendez 2004; Socas-Navarro \& Norton 2007;
Caffau et al. 2008; but see Ayres et al. 2006, Ayres 2008, and 
Centeno \& Socas-Navarro 2008 for a departing conclusion).
This large revision in the oxygen abundance has an important impact in many areas of 
astrophysics. In particular, solar models computed with the low oxygen abundance 
do not agree with precise helioseismological measurements (see Basu \& Antia 2008
and references therein). 

If the revised solar oxygen abundance is correct,
this would imply that the standard solar structure model may be missing (or have a too 
simplified treatment of) important physical processes, which would carry over to
errors in the calculations of stellar evolution at large.
On the other hand, if the fault lies with the new solar photospheric abundance
analyses it raises concerns about how well we understand stellar atmospheres and
spectral line formation in general with wider implications for  
Galactic chemical evolution studies relying on accurate stellar abundance measurements.
Indeed, the oxygen content in metal-poor stars is currently under debate
(see Asplund \& Garc{\'{\i}}a P{\'e}rez 2001; Mel\'endez et al. 2001, 2006, and 
references therein).

Since the Sun is used as a fundamental reference in most areas of astrophysics,
it is important to study all spectral features available
for abundance analysis. Recent studies have mainly used the  [\ion{O}{i}]  6300, 6363 \AA\
forbidden lines, the \ion{O}{i} 7777 \AA\ triplet, pure rotation OH lines
as well as the fundamental vibration-rotation OH lines 
(Allende Prieto et al. 2001, 2004; Asplund et al. 2004),
from which $\log \epsilon_{\rm O} = 8.66 \pm 0.05$ have been recommended (Asplund et al. 2004).
Mel\'endez (2004) presented a study of the first-overtone OH lines, which are
unfortunately very weak for a precise determination of the oxygen abundance,
but within the uncertainties these lines also support a low 
solar oxygen abundance.

Another independent way to estimate the solar oxygen abundance is using
the forbidden line at 5577.34 \AA. This line has been disregarded in the past
due to blending with C$_2$ lines (e.g. Altrock 1968; Lambert 1978).
Here, we present the first detailed study of the [\ion{O}{i}] 5577 \AA\ line,
carefully taking into account the blends by C$_2$ lines.
The advantage of this feature is that the C$_2$ blends introduce a large asymmetry 
in the profile, allowing thus to readily constrain their contribution.
Another advantage is that there are other nearby C$_2$ lines that can be used 
to calibrate the blending C$_2$ lines, and therefore to determine a 
reliable solar oxygen abundance.

\begin{figure}
%\centering
\includegraphics[width=\hsize]{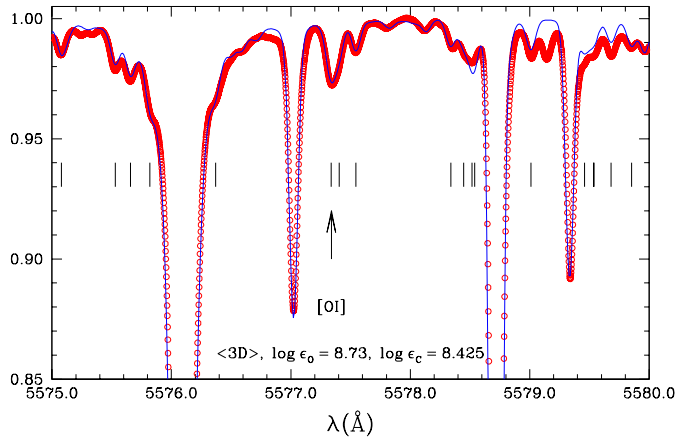}
\caption{Observed solar spectrum (Delbouille et al. 1973) between 5575 and 5580\,\AA\
(red circles) and the best fit employing the $<$3D$>$ model (solid blue line).
The locations of the most prominent C$_2$ lines used to estimate the C abundance
are denoted with vertical lines, while the position of the [\ion{O}{i}] line is marked with an arrow.
The resulting solar C abundance is  $\log \epsilon_{\rm C} = 8.42$ with which the 
5577.3\,\AA\ feature implies a solar O abundance of $\log \epsilon_{\rm O} = 8.73$.
}
\label{f:5575-5580}
 \end{figure}

\begin{figure}
%\centering
\includegraphics[width=\hsize]{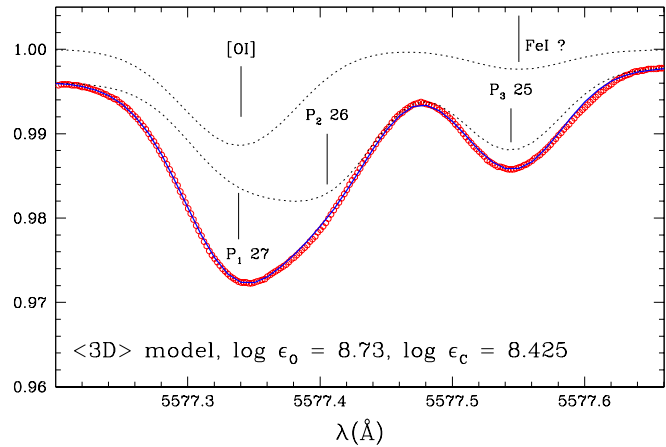}
\includegraphics[width=\hsize]{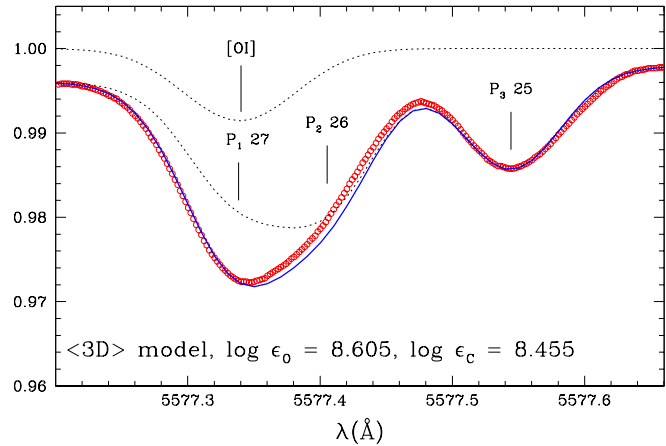}
\caption{{\em Upper panel:} An enlargement of the observed (red circles, Delbouille et al. 1973) 
and $<$3D$>$-based theoretical profiles (solid blue line) shown in Fig. \ref{f:5575-5580}. 
The dotted lines show the separate contribution of [\ion{O}{i}] and other lines labeled 
accordingly. 
The 5577.54 {\AA} P$_3$25 C$_2$ line is blended with an unknown feature, assumed here
to be \ion{Fe}{I}. 
{\em Lower panel:} Same as above but assuming instead that the red feature is entirely due to
P$_3$25 C$_2$. Applying this 0.03\,dex higher C abundance to the [\ion{O}{i}]+C$_2$ blend 
implies a smaller left-over contribution for [\ion{O}{i}] and thus a 0.13\,dex lower O abundance
at the expense of a significantly poorer profile fit. 
}
\label{f:oi}
 \end{figure}

\section{Analysis}

We used the Delbouille et al. (1973) disk-center solar spectrum 
(often called the Liege solar atlas),
the Brault \& Neckel (1987) disk-center intensity atlas (see Neckel 1999),
and the Kurucz et al. (1984) solar-flux spectrum.
These solar spectra are of extremely high resolution
(resolving power R $\approx$ 500 000 - 700 000)
and have a S/N of several thousands at 5577 \AA.
The solar oxygen abundance obtained with these different atlases
agree within $\sigma$ = 0.04\,dex. 
In Fig. 1 the Delbouille et al. (1973) solar spectrum
around the 5577.3 \AA\ [\ion{O}{i}] line is shown as open circles. 

The oscillator strength for the [\ion{O}{i}] 5577 \AA\ line
is well determined:
the mean of the available calculations is 
$\log gf = -8.25 \pm 0.03$ (Galavis et al. 1997;
Baluja \& Zeippen 1988; Fischer \& Saha 1983; Nicolaides \&  Sinano{\u g}lu 1973).
We adopted  $\log gf  = -8.28$ from  Galavis et al. (1997),
which is on the same scale as the log {\it gf}
value adopted for the 6300, 6363 \AA\ forbidden lines (Asplund et al. 2004) within
0.007\,dex;
had we instead opted for the mean value of the above-mentioned 
theoretical $gf$-values our derived O abundance would
be 0.03\,dex lower. The excitation potential of the [\ion{O}{i}] 5577 \AA\ line
is 1.967 eV.
According to the NIST database\footnote{http://physics.nist.gov/PhysRefData/ASD/index.html}
the observed and Ritz wavelengths are 5577.34 and 5577.339 \AA, respectively. 
Small errors in the wavelength translate into only small uncertainties
(at the level of 0.01 dex) in the O abundances.

We included in the spectral synthesis other atomic and molecular
(C$_2$ Swan and CN red systems) lines present around  5577 \AA, 
but the main contributors to the 5577.3 \AA\ feature are mostly three lines:
the [\ion{O}{i}] line and the P$_1$27 and P$_2$26 lines of the C$_2$ (1-2) band.
Atomic lines were taken from the latest Kurucz line lists\footnote{http://kurucz.harvard.edu/} 
and the CN line list described in Mel\'endez \& Barbuy (1999) was adopted.
The line list for the 0-1 and 1-2 bands of the C$_2$ Swan system
was constructed in a similar way to the 0-0 band described in
Mel\'endez \& Cohen (2007), but with the line positions taken primarily from 
Tanabashi et al. (2007) and complemented with Phillips \& Davis (1968) for
the higher excitation lines not included by Tanahashi et al.. 
Relative oscillator strengths for the 0-1 and 1-2 bands were taken from Kokkin et al. (2007),
normalizing those values to the laboratory oscillator strength of the 0-0 band
recommended by Grevesse et al. (1991), $f_{00}$ = 3.03 $\times 10^{-2}$.
The rotational strengths were computed following Kovacs (1961) and the
rotational dependence of the band strengths was taken into
account (Dwivedi et al. 1978).

A set of six different model atmospheres was employed:
a three-dimensional hydrodynamical model of the solar atmosphere 
(Asplund et al. 2000; here denoted 3D) 
and its temporal and spatial average (here: $<$3D$>$)
(Asplund et al. 2004), MARCS model (Asplund et al. 1997),
Kurucz overshooting (Castelli et al. 1997) and the latest no-overshooting 
(Castelli \& Kurucz 2004) models and the semi-empirical Howeger-M\"uller (1974) model.
The spectral line formation was calculated in 3D and 1D using the same spectrum
synthesis code as
described in Asplund et al. (2004). The 1D calculations were
also performed using the 2002 version of MOOG 
(Sneden 1973); the different 1D calculations
with different codes agree at the level of 0.01\,dex.

The contribution from the two C$_2$ lines (P$_1$27 and P$_2$26) to the 5577.3\,\AA\
feature can be well constrained from 
the red asymmetry in the profile. In addition there are many neighboring C$_2$ lines of very similar
excitation potential and line strengths and thus basically identical line formation. 
Note that the error in the relative
strengths of the C$_2$ lines is  negligible, as the rotational strengths 
are precisely given by their H\"onl-London factors (Kovacs 1961).
Fig. \ref{f:5575-5580} shows the best fit to the 5575-5580\,\AA\ region of the Liege disk-center
atlas based on the 3D-averaged model atmosphere; the locations of the most important
C$_2$ lines are marked with vertical lines. Clearly the overall agreement is quite
satisfactory. Also encouraging is that the thus estimated solar C abundance  is 
as expected for the different model atmospheres (Asplund et al. 2005).
As seen in Fig. \ref{f:oi}, the profile of the [\ion{O}{i}]+C$_2$ feature is very well 
described with this C abundance and a solar oxygen abundance of 
$\log \epsilon_{\rm O} = 8.73$\footnote{Note that the abundance quoted here is only
from the Liege intensity atlas (Figs. 1 and 2). The final abundance given in Table 1 is based on 
the mean from the three different solar atlases.} 
with the $<$3D$>$ model. 
The resulting disk-center intensity equivalent width for only the [\ion{O}{i}] line is 1.23\,m\AA ,
which corresponds to $\log \epsilon_{\rm O} = 8.72$ for the full 3D model.
The Kurucz et al. (1984) solar-flux and the Brault \& Neckel (1987) intensity spectrum gave an O abundance 
0.01 dex higher and -0.07 dex lower, respectively.
Results for different model atmospheres are similar, as shown in Table \ref{t:oxygen},
thus the [\ion{O}{i}] 5577 \AA\ line is almost insensitive to the adopted model atmosphere.

It should be noted that the predicted profile for the 
neighboring feature at 5577.55 \AA\ is too weak with the C abundance from
the overall fit to all C$_2$ lines in the wavelength window 5575-5580\,\AA\
(Fig. \ref{f:oi}), suggesting that it is blended. Scouring available atomic databases did
not reveal any likely candidates. We have therefore tentatively assigned the blending
line as being due to \ion{Fe}{i} but the exact choice is unimportant.
Relying solely on the P$_3$25  transition and without accounting for this blend, 
the estimated C abundance would have been 0.03\,dex higher.
Because of the larger contribution from C$_2$ the [\ion{O}{i}] part of the 
5577.3\,\AA\ feature must decrease correspondingly, leading to a much reduced
solar O abundance: $\log \epsilon_{\rm O} \approx 8.6$.
In this case, however, the profile fit is very unsatisfactory, in particular in 
the red wing, as evident in Fig. \ref{f:oi}. 

The main errors are the uncertainties in the log {\it gf} value of the [\ion{O}{i}] line
(assumed here to be 0.03\,dex, which is the $\sigma$ between the different calculations;
c.f. Galavis et al. 1997; Baluja \& Zeippen 1988; Fischer \& Saha 1983; Nicolaides \&  Sinano{\u g}lu 1973)
and the observational error. Part of the observational error (0.04 dex) may be due 
to variation with solar activity (e.g. Livingston et al. 2007), and it was estimated by 
using different solar atlases. Note that Caffau et al. (2008) has also found a similar
scatter (0.03 dex) between the [\ion{O}{i}] 6300\AA\ oxygen abundances obtained from different
solar atlases. The other part of the observational error (0.05 dex) was estimated
by employing different assumptions with respect to the continuum determination 
and varying by 0.01 \AA\ the wavelengths of the C$_2$ lines\footnote{According to
Tanabashi et al. (2007) the line positions for those C$_2$ lines are not very
precise, hence we allowed small changes by up to 0.01 \AA\ in wavelength.
Also, note that Allende Prieto \& Garc\'ia L\'opez (1998) have shown that the Delbouille et al.
solar disk-center intensity atlas have slight errors in its absolute wavelength calibration,
although probably not over a wavelength scale of a few \AA , which is relevant here.}
and of the unknown feature at 5577.55 \AA.
Given the excellent agreement between the oxygen abundance obtained from different model atmospheres 
for this spectral feature, we estimate that the errors introduced by
problems in the model atmosphere should be 0.04\,dex at most.
Non-LTE effects for this [\ion{O}{i}] line should be negligible (e.g. Altrock 1968; Takeda 1994).
Adding the squared errors (0.03 dex in the log {\it gf} value,
0.04 dex in the apparent line variation, 0.05 dex due to uncertainties in the continuum
and line positions, and 0.04 dex in the models), we estimate a conservative total error of 0.08\,dex.

\begin{table*}
\begin{minipage}[t]{\textwidth}
\caption{Solar oxygen abundance from different model atmospheres.}
\label{t:oxygen}
\centering          
\renewcommand{\footnoterule}{}  % to avoid a line before footnotes
\begin{tabular}{llllllll} 
\hline\hline                
lines & 3D & $<$3D$>$ &  Holweger-M\"uller & MARCS & Kurucz overshooting & Kurucz no overshooting \\
\hline    
{[\ion{O}{i}]} 5577 \AA \footnote{Average abundance from the Delbouille et al. (0.02 higher), 
Brault \& Neckel (0.04 lower) and Kurucz et al. (0.03 higher) atlases.}
                       & 8.70$\pm$0.08 & 8.71 $\pm$ 0.08 & 8.73$\pm$0.08 & 8.70$\pm$ 0.08 & 8.74 $\pm$ 0.08 & 8.70 $\pm$ 0.08\\
\\ 
{[\ion{O}{i}]}         & 8.69 $\pm$ 0.05 & 8.73 $\pm$ 0.05 & 8.75 $\pm$ 0.05 & 8.71 $\pm$ 0.05 & 8.77 $\pm$ 0.05 & 8.73 $\pm$ 0.05\\ 
\\
 \ion{O}{i}            & 8.67 $\pm$ 0.05 & 8.68 $\pm$ 0.05 & 8.64 $\pm$ 0.08 & 8.72 $\pm$ 0.05 & 8.67 $\pm$ 0.05 & 8.64 $\pm$ 0.05\\
 \\
OH                     & 8.61 $\pm$ 0.10 & 8.68 $\pm$ 0.10 & 8.85 $\pm$ 0.10 & 8.73 $\pm$ 0.11 & 8.80 $\pm$ 0.10 & 8.72 $\pm$ 0.10\\
\\
Average               & 8.67 $\pm$ 0.04 & 8.70 $\pm$ 0.04 & 8.71 $\pm$ 0.10 & 8.72 $\pm$ 0.04 &  8.73 $\pm$ 0.07 & 8.69 $\pm$ 0.05\\
\hline                                 
\end{tabular}
\end{minipage}
\end{table*}

\section{Solar oxygen abundance from different diagnostics}

Asplund et al. (2004) have determined the solar oxygen abundance from
the [\ion{O}{i}] 6300, 6363 \AA\ forbidden lines, \ion{O}{i} permitted lines, and the pure rotation and
fundamental vibration-rotation OH lines, employing a 3D hydrodynamical simulation of the
solar atmosphere as well as a 1D MARCS 
(Asplund et al. 1997) and the Holweger \& M\"uller (1974) semi-empirical model atmospheres.
Mel\'endez (2004) has added oxygen abundances determined from
the first-overtone OH lines, as well as computed the solar oxygen abundance 
for the above features using the $<$3D$>$ and Kurucz (Castelli et al. 1997) convective 
overshooting models. Here we add the 5577.3 \AA\ forbidden line for the five model 
atmospheres described above, and we also have determined oxygen abundances 
for the latest Kurucz (Castelli \& Kurucz 2004) model without convective overshooting.

The results of the different computations are summarized in Table 1,
where the mean oxygen abundance for the forbidden lines now includes
the 5577.3 \AA\ line. 
The mean value given for the infrared OH lines includes all the
OH mentioned above, but given half-weight to the pure rotation and the
first-overtone line, because the pure rotation lines are quite
sensitive to the detailed structure of the model atmosphere, while the first-overtone
lines are very weak and the observational error is high.
For the \ion{O}{i} lines we adopt the mean non-LTE results of Allende Prieto et al. (2004) 
and Asplund et al. (2004), with and without including inelastic H collisions, respectively.
As discussed by Asplund (2005), the H collisions are based on the classical but 
highly uncertain formula by Drawin (1968), yet the O abundance with (and without)
collisions is only $\approx$ 0.03 higher (lower) than the mean O abundance adopted here.
We also show in the last row the averaged solar oxygen abundance
for each model atmosphere. In obtaining this mean value, half weight
has been given to the result obtained from the molecular lines, because
of their strong model atmosphere sensitivity.

As can be seen, all model atmospheres favor a
low solar oxygen abundance: $\log \epsilon_{\rm O} \simeq 8.7$. The reduction in
the solar oxygen abundance from the historical value of $\simeq 8.9$ 
(Anders \& Grevesse 1989)\footnote{It should be noted that the solar O abundance of 8.83 given
in the compilation of Grevesse \& Sauval (1998) is only a preliminary analysis based on a 
Holweger-M\"uller (1974) model with a modified temperature structure to remove various 
abundance trends with excitation potential and line strengths for species such 
as \ion{Fe}{i}, OH and CH.} to the present level is therefore not primarily driven by the
use of a 3D hydrodynamic models. In this context,  improvements in the input
atomic and molecular data, more realistic non-LTE calculations and a 
more careful treatment of blends are equally important.
The average O abundance for the 3D model presented here is 8.67, only 0.01 higher than
the value recommended by Asplund et al. (2004, $\log \epsilon_{\rm O} = 8.66 \pm 0.05$).
Note that the recommended O abundance obtained with the Holweger \& M\"uller (1974) model 
is only 0.04\,dex higher than the 3D-based abundance but show the largest scatter (0.10\,dex),
mainly due to the large difference between OH and \ion{O}{i}. 

Using 1/$\sigma^2$ of the mean oxygen abundances given
in Table \ref{t:oxygen} as weights (assuming a minimum $\sigma = 0.04$\,dex) implies
a weighted mean solar oxygen abundance from the six solar model atmospheres of  
$\log \epsilon_{\rm O} = 8.70$ ($\sigma = 0.02$).

Higher solar O abundances have recently been advocated by Caffau et al. (2008)
and Ayres (2008) based on the atomic lines and an alternative 3D solar model computed with
the CO5BOLD code (Freytag et al. 2002); neither study includes the here employed
[\ion{O}{i}] 5577\,\AA\ line. Caffau et al. (2008) finds $\log \epsilon_{\rm O} = 8.76$ with the
main differences being the adopted equivalent widths and the less severe non-LTE abundance
corrections for \ion{O}{i} due to their inclusion of inelastic H collisions close to the 
Drawin (1968) recipe; it appears that the results of Asplund et al. (2004) and 
Caffau et al. (2008) in most cases are in excellent agreement when otherwise
using the same line input data, in spite of some differences in the mean 
temperature stratification of the two 3D models.
Ayres (2008) found $\log \epsilon_{\rm O} = 8.81$ based on the [\ion{O}{i}] 6300\,\AA\ line,
which is blended by a \ion{Ni}{i} line (Allende Prieto et al. 2001).
In order to achieve the best overall profile fit, he had to use a solar Ni abundance
lower by 0.15\,dex than the value given by Grevesse et al. (2007). 
This is well outside the quoted uncertainties in the laboratory $gf$-value 
(Johansson et al. 2003) and the solar Ni abundance. 
Clearly more work is needed in order to resolve the existing issues with solar
O abundance determinations using the different indicators. 
We are currently working towards this goal using a new, further improved
3D solar model with a more realistic radiative transfer treatment (e.g. opacity sampling
instead of opacity binning).

\section{Conclusions}

A low oxygen abundance is obtained from the [\ion{O}{i}] 5577.3 \AA\ line, 
almost independent of the adopted model atmosphere: 
$\log \epsilon_{\rm O} = 8.71 \pm 0.02 \pm 0.07$ (0.02 dex is the $\sigma$ from 
using different models, and 0.07 dex is the error due to uncertainties in
the the log $gf$ value (0.03 dex), apparent line variation (0.04 dex) and uncertainties 
in the continuum and line positions (0.05 dex)).
This value is close to the results of Asplund et al. (2004) and Mel\'endez (2004) 
for other solar oxygen abundance indicators 
(\ion{O}{i}, [\ion{O}{i}] and OH lines). Including those abundances and 
employing six different 3D and 1D model atmospheres, 
we estimate a solar O abundance of $\log \epsilon_{\rm O} = 8.70$ ($\sigma$ = 0.02)
with only a small sensitivity to the employed model atmosphere. 

The low solar oxygen abundance that we propose here
as well as the downward revision of other solar chemical abundances 
(Asplund  2000, 2005, 2007; Asplund et al. 2000, 2005, 2006)
pose a challenge to standard solar models in light of helioseismological 
observations (e.g. Antia \& Basu 2005, 2006; 
Bahcall et al. 2004, 2005, 2006; Basu \& Antia 2004, 2008; Chaplin et al. 2007;
Guzik et al. 2005; Yang \& Bi 2007).

\begin{acknowledgements}
We thank A. Alves-Brito \& B. Barbuy for sending a copy of the
Phillips \& Davis (1968) Berkeley atlas of the C$_2$ Swan system.
This work has been supported by ARC (DP0588836) and 
FCT (project PTDC/CTE-AST/65971/2006).
\end{acknowledgements}


\begin{thebibliography}{}
\bibitem[Allende Prieto 
\& Garcia Lopez(1998)]{1998A&AS..129...41A} Allende Prieto, C., \& Garcia Lopez, R.~J.\ 1998, \aaps, 129, 41 

\bibitem[Allende Prieto et al.(2001)]{2001ApJ...556L..63A} Allende Prieto, 
C., Lambert, D.~L., \& Asplund, M.\ 2001, \apjl, 556, L63 

\bibitem[Allende Prieto et 
al.(2004)]{2004A&A...423.1109A} Allende Prieto, C., Asplund, M., \& Fabiani Bendicho, P.\ 2004, \aap, 423, 1109 

\bibitem[Altrock(1968)]{1968SoPh....5..260A} Altrock, R.~C.\ 1968, 
\solphys, 5, 260 

\bibitem[Anders 
\& Grevesse(1989)]{1989GeCoA..53..197A} Anders, E., \& Grevesse, N.\ 1989, \gca, 53, 197 

\bibitem[Antia 
\& Basu(2005)]{2005ApJ...620L.129A} Antia, H.~M., \& Basu, S.\ 2005, \apjl, 620, L129 

\bibitem[Antia 
\& Basu(2006)]{2006ApJ...644.1292A} Antia, H.~M., \& Basu, S.\ 2006, \apj, 644, 1292 

\bibitem[Asplund et 
al.(1997)]{1997A&A...318..521A} Asplund, M., Gustafsson, B., Kiselman, D., \& Eriksson, K.\ 1997, \aap, 318, 521 

\bibitem[Asplund(2000)]{2000A&A...359..755A} Asplund, M.\ 2000, \aap, 359, 755 

\bibitem[Asplund et 
al.(2000)]{2000A&A...359..743A} Asplund, M., Nordlund, {\AA}., Trampedach, R., \& Stein, R.~F.\ 2000, \aap, 359, 743 

\bibitem[Asplund 
\& Garc{\'{\i}}a P{\'e}rez(2001)]{2001A&A...372..601A} Asplund, M., \& Garc{\'{\i}}a P{\'e}rez, A.~E.\ 2001, \aap, 372, 601 


\bibitem[{{Asplund} {et~al.}(2004){Asplund}, {Grevesse}, {Sauval}, {Allende
  Prieto}, \& {Kiselman}}]{2004A&A...417..751A}
{Asplund}, M., {Grevesse}, N., {Sauval}, A.~J., {Allende Prieto}, C., \&
  {Kiselman}, D. 2004, \aap, 417, 751

\bibitem[{{Asplund}(2005)}]{2005ARA&A..43..481A}
{Asplund}, M. 2005, \araa, 43, 481

\bibitem[Asplund et 
al.(2005)]{2005A&A...431..693A} Asplund, M., Grevesse, N., Sauval, A.~J., Allende Prieto, C., \& Blomme, R.\ 2005, \aap, 431, 693 

\bibitem[Asplund et al.(2006)]{2006CoAst.147...76A} Asplund, M., Grevesse, 
N., \& Sauval, A.~J.\ 2006, Communications in Asteroseismology, 147, 76 

\bibitem[Asplund(2007)]{2007IAUS..239..122A} Asplund, M.\ 2007, IAU 
Symposium, 239, 122 

\bibitem[Ayres et al.(2006)]{2006ApJS..165..618A} Ayres, T.~R., Plymate, 
C., \& Keller, C.~U.\ 2006, \apjs, 165, 618 

\bibitem[Ayres (2008)]{} Ayres, T.~R.\ 2008, \apjs, submitted 

\bibitem[Bahcall et al.(2004)]{2004ApJ...614..464B} Bahcall, J.~N., 
Serenelli, A.~M., \& Pinsonneault, M.\ 2004, \apj, 614, 464 

\bibitem[Bahcall et al.(2005)]{2005ApJ...631.1281B} Bahcall, J.~N., Basu, 
S., \& Serenelli, A.~M.\ 2005, \apj, 631, 1281 

\bibitem[Bahcall et al.(2006)]{2006ApJS..165..400B} Bahcall, J.~N., 
Serenelli, A.~M., \& Basu, S.\ 2006, \apjs, 165, 400 

\bibitem[Baluja 
\& Zeippen(1988)]{1988JPhB...21.1455B} Baluja, K.~L., \& Zeippen, C.~J.\ 1988, Journal of Physics B Atomic Molecular Physics, 21, 1455 


\bibitem[Brault 
\& Neckel(1987)]{} Brault, J., \& Neckel, H. 1987, Spectral atlas of solar absolute 
disk-averaged and disk-center intensity from 3290 to 12510 \AA

\bibitem[Basu 
\& Antia(2004)]{2004ApJ...606L..85B} Basu, S., \& Antia, H.~M.\ 2004, \apjl, 606, L85 

\bibitem[Basu 
\& Antia(2008)]{2008PhR...457..217B} Basu, S., \& Antia, H.~M.\ 2008, \physrep, 457, 217 

\bibitem[Caffau et al.(2008)]{2008arXiv0805.4398C} Caffau, E., Ludwig, 
H.-G., Steffen, M., Ayres, T.~R., Bonifacio, P., Cayrel, R., Freytag, B., 
\& Plez, B.\ 2008, \aap, in press (arXiv:0805.4398)

\bibitem[Castelli et 
al.(1997)]{1997A&A...318..841C} Castelli, F., Gratton, R.~G., \& Kurucz, R.~L.\ 1997, \aap, 318, 841 

\bibitem[Castelli 
\& Kurucz(2004)]{2004A&A...419..725C} Castelli, F., \& Kurucz, R.~L.\ 2004, \aap, 419, 725 

\bibitem[Centeno 
\& Socas-Navarro(2008)]{2008arXiv0803.0990C} Centeno, R., \& Socas-Navarro, H.\ 2008, \apjl, in press (arXiv:0803.0990)


\bibitem[Chaplin et al.(2007)]{2007ApJ...670..872C} Chaplin, W.~J., 
Serenelli, A.~M., Basu, S., Elsworth, Y., New, R., 
\& Verner, G.~A.\ 2007, \apj, 670, 872 

\bibitem[Delbouille et al.(1973)]{1973apds.book.....D} Delbouille, L., 
Roland, G., 
\& Neven, L.\ 1973, Liege: Universite de Liege, Institut d'Astrophysique, 1973 

\bibitem[Drawin(1968)]{1968ZPhy..211..404D} Drawin, H.-W.\ 1968, 
Zeitschrift fur Physik , 211, 404 

\bibitem[Dwivedi et al.(1978)]{1978ApJS...36..573D} Dwivedi, P.~H., Branch, 
D., Huffaker, J.~N., \& Bell, R.~A.\ 1978, \apjs, 36, 573 

 \bibitem[Fischer 
\& Saha(1983)]{1983PhRvA..28.3169F} Fischer, C.~F., \& Saha, H.~P.\ 1983, \pra, 28, 3169 

\bibitem[Freytag et al.(2002)]{2002AN....323..213F} Freytag, B., Steffen, 
M., \& Dorch, B.\ 2002, Astronomische Nachrichten, 323, 213 

\bibitem[Galavis et 
al.(1997)]{1997A&AS..123..159G} Galavis, M.~E., Mendoza, C., \& Zeippen, C.~J.\ 1997, \aaps, 123, 159 

\bibitem[Grevesse et 
al.(1984)]{1984A&A...141...10G} Grevesse, N., Sauval, A.~J., \& van Dishoeck, E.~F.\ 1984, \aap, 141, 10 

\bibitem[Grevesse et 
al.(1991)]{1991A&A...242..488G} Grevesse, N., Lambert, D.~L., Sauval, A.~J., van Dishoek, E.~F., Farmer, C.~B., \& Norton, R.~H.\ 1991, \aap, 242, 488 

\bibitem[Grevesse 
\& Sauval(1998)]{1998SSRv...85..161G} Grevesse, N., \& Sauval, A.~J.\ 1998, Space Science Reviews, 85, 161 

\bibitem[Grevesse et al.(2007)]{2007SSRv..130..105G} Grevesse, N., Asplund, 
M., \& Sauval, A.~J.\ 2007, Space Science Reviews, 130, 105 

\bibitem[Guzik et al.(2005)]{2005ApJ...627.1049G} Guzik, J.~A., Watson, 
L.~S., \& Cox, A.~N.\ 2005, \apj, 627, 1049 

\bibitem[Holweger 
\& M\"uller(1974)]{1974SoPh...39...19H} Holweger, H., \& M\"uller, E.~A.\ 1974, \solphys, 39, 19 


\bibitem[Holweger(2001)]{2001AIPC..598...23H} Holweger, H.\ 2001, Joint 
SOHO/ACE workshop ''Solar and Galactic Composition'', 598, 23 

\bibitem[Johansson et al.(2003)]{2003ApJ...584L.107J} Johansson, S., 
Litz{\'e}n, U., Lundberg, H., \& Zhang, Z.\ 2003, \apjl, 584, L107 


\bibitem[Kokkin et al.(2007)]{2007JChPh.126h4302K} Kokkin, D.~L., Bacskay, 
G.~B., \& Schmidt, T.~W.\ 2007, \jcp, 126, 4302 

\bibitem[Kovacs(1961)]{} Kovacs, I. 1961, Rotational structure in the spectra of Diatomic
Molecules (London: Hilger)

\bibitem[Kurucz et al.(1984)]{1984sfat.book.....K} Kurucz, R.~L., Furenlid, 
I., Brault, J., 
\& Testerman, L.\ 1984, National Solar Observatory Atlas, Sunspot, New Mexico: National Solar Observatory, 1984,  

\bibitem[Lambert(1978)]{1978MNRAS.182..249L} Lambert, D.~L.\ 1978, \mnras, 
182, 249 


\bibitem[Livingston et al.(2007)]{2007ApJ...657.1137L} Livingston, W., 
Wallace, L., White, O.~R., \& Giampapa, M.~S.\ 2007, \apj, 657, 1137 


\bibitem[Mel{\'e}ndez 
\& Barbuy(1999)]{1999ApJS..124..527M} Mel{\'e}ndez, J., \& Barbuy, B.\ 1999, \apjs, 124, 527 

\bibitem[Mel{\'e}ndez et al.(2001)]{2001ApJ...556..858M} Mel{\'e}ndez, J., 
Barbuy, B., \& Spite, F.\ 2001, \apj, 556, 858 

\bibitem[Mel{\'e}ndez(2004)]{2004ApJ...615.1042M} Mel{\'e}ndez, J.\ 2004, 
\apj, 615, 1042 

\bibitem[Mel{\'e}ndez et al.(2006)]{2006ApJ...642.1082M} Mel{\'e}ndez, J., 
Shchukina, N.~G., Vasiljeva, I.~E., \& Ram{\'{\i}}rez, I.\ 2006, \apj, 642, 1082 


\bibitem[Mel{\'e}ndez 
\& Cohen(2007)]{2007ApJ...659L..25M} Mel{\'e}ndez, J., \& Cohen, J.~G.\ 2007, \apjl, 659, L25 


\bibitem[Neckel, H.]{1999SoPh...184...421N} Neckel, H.\ 1999, \solphys, 184, 421 


\bibitem[Nicolaides 
\& Sinano{\u g}lu(1973)]{1973SoPh...29...17N} Nicolaides, C.~A., \& Sinano{\u g}lu, O.\ 1973, \solphys, 29, 17 


\bibitem[Phillips \& Davis(1968)]{1968sscm.book.....P} Phillips, J.~G., \& Davis, S.~P.\ 1968, Berkeley Analyses of Molecular Spectra, Berkeley: University of California Press, 1968,  

%\bibitem[{{Ram{\'{\i}}rez} {et~al.}(2007){Ram{\'{\i}}rez}, {Allende Prieto}, \&
%  {Lambert}}]{2007A&A...465..271R}
%{Ram{\'{\i}}rez}, I., {Allende Prieto}, C., \& {Lambert}, D.~L. 2007, \aap,
%  465, 271

\bibitem[Sauval et al.(1984)]{1984ApJ...282..330S} Sauval, A.~J., Grevesse, 
N., Zander, R., Brault, J.~W., \& Stokes, G.~M.\ 1984, \apj, 282, 330 

\bibitem[{{Sneden}(1973)}]{1973PhDT.......180S}
{Sneden}, C.~A. 1973, PhD thesis, AA(THE UNIVERSITY OF TEXAS AT AUSTIN.)

\bibitem[Socas-Navarro 
\& Norton(2007)]{2007ApJ...660L.153S} Socas-Navarro, H., \& Norton, A.~A.\ 2007, \apjl, 660, L153 

\bibitem[Takeda(1994)]{1994PASJ...46...53T} Takeda, Y.\ 1994, \pasj, 46, 53 

\bibitem[Tanabashi et al.(2007)]{2007ApJS..169..472T} Tanabashi, A., Hirao, 
T., Amano, T., \& Bernath, P.~F.\ 2007, \apjs, 169, 472 

\bibitem[Yang 
\& Bi(2007)]{2007ApJ...658L..67Y} Yang, W.~M., \& Bi, S.~L.\ 2007, \apjl, 658, L67 


\end{thebibliography}
\end{document}